\documentclass[%
 reprint,
 amsmath,amssymb,
 aps,
]{revtex4-1}
\usepackage[utf8x]{inputenc}
\usepackage{braket}
\usepackage{hyperref}
\usepackage{amsmath}
\usepackage{bbold}
\usepackage{amssymb}
\usepackage{graphicx}
\usepackage{dcolumn}
\usepackage{bm}

\begin{document}

\preprint{APS/123-QED}

\title{Expected Optimal Time for the NMR Implementation of Shor’s Algorithm for Factorising 15}

\author{Vlad C\u{a}rare}
 \email{v.carare@lancaster.ac.uk}
\author{Alejandro Cros Carrillo de Albornoz}%
 \email{a.cros-carrillo-de-albornoz@sms.ed.ac.uk}
 \author{John Taylor}
 \email{j.t.taylor@lancaster.ac.uk}
\affiliation{%
Lancaster University, United Kingdom\\
}%

\date{\today}

\begin{abstract}
In this paper, we briefly discuss the methodology for simulating a quantum computer which performs
Shor’s algorithm on a 7-qubit system to factorise 15. Using this simulation and the overlooked quantum brachistochrone method, we devised a Monte Carlo algorithm to estimate the expected time a theoretical quantum computer could
perform this calculation under the same energy conditions as current working
quantum computers. We found that, experimentally, a nuclear magnetic resonance quantum computer
would take $1.59 \pm 0.04 \text{s}$ to perform our simulated computation, whereas the expected optimal time under the same energy conditions is $0.955 \pm 0.004$ ms.
Moreover, we found that the expected time is inversely proportional to the energy
variance of our qubit states (as expected). Finally, we propose this theoretical method for analysing the time-efficiency of future quantum computing experiments.

\end{abstract}

\maketitle


\section{\label{Sec:Introduction}Introduction}

\quad It is trivial for a classical computer to multiply two large prime numbers. However, it is difficult to do the opposite: to find the prime factors of an extremely large number. Common forms of cryptography such as the RSA algorithm \cite{Rivest} use arbitrarily large prime numbers as keys to encrypt messages. Without knowing any of these prime factors, classical computers can only decrypt these messages via a process of trial and error. For numbers that are hundreds of digits long, even the most powerful computers today would take hundreds of years to process the factorisation \cite{BruteForce}. Quantum computers provide a more sophisticated way of decoding such messages, through Shor's algorithm \cite{Shor}. A physical implementation of Shor's algorithm for factorising 15 has been realised using a nuclear magnetic resonance scheme \cite{Vandersypen}. 

There is, however, scope for minimisation in the operation time. Using the overlooked theory of the quantum brachistrochrone, laid out by Carlini et al. in their seminal paper \cite{Carlini}, we can calculate a theoretical optimal time for this implementation. The application of the quantum brachistochrone theory to Shor's algorithm for factorising 15 is, to the best of our knowledge, not available in literature.

We devised a new method based on a Monte Carlo simulation to find the expected optimal time of the gates used, given the experimental constraints of \cite{Vandersypen}. Further, we compared our result to the experimental duration that we calculated using reported values from \cite{Vandersypen}. Even though our paper is based on this experiment, the method that we implement need not be. Thus, there is scope for extension to other algorithms and experiments.

Section \ref{Sec:Shor'sAlgorithm} gives a summary of Shor's factorisation algorithm.

Section \ref{Sec:TimeOptimalSection} covers the fundamentals of quantum brachistochrone theory.

Section \ref{Sec:ExperimentalComparison} looks at the nuclear magnetic resonance implementation of Shor's algorithm for factorising the number 15.

Section \ref{Sec:TimeEfficiencyOfExperiment} builds on the previous two sections and details the method that we used to find out the expected optimal time for Shor's algorithm for factorising 15. 

Section \ref{Sec:Results & Conclusion} offers a comparison between the experimental duration and the expected optimal time of Shor's algorithm for factorising 15.

\section{\label{Sec:Shor'sAlgorithm}Shor's Factorisation Algorithm}

\begin{figure*}[t]
	\begin{center}
	\includegraphics[width=15cm]{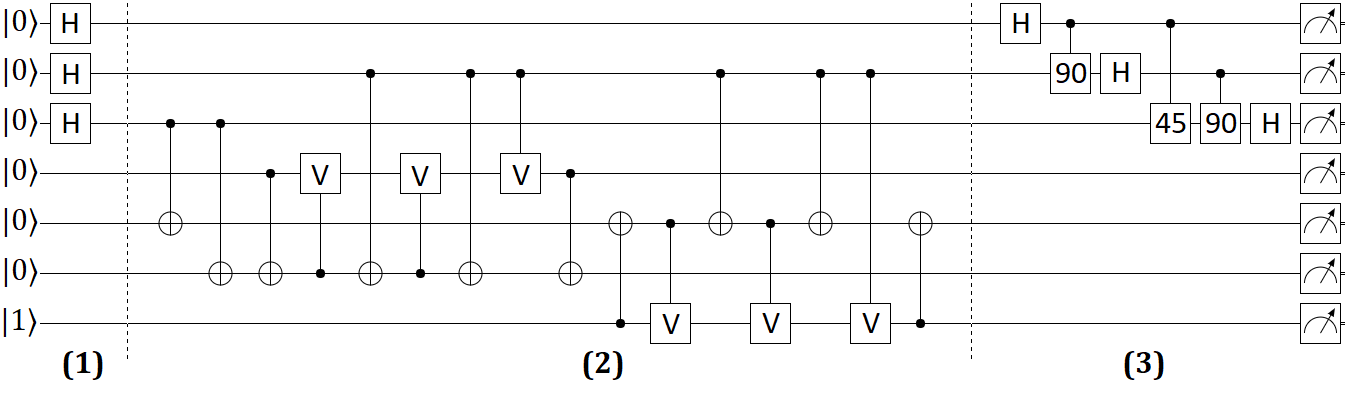}
	\caption{A gate by gate breakdown of the 7-qubit Shor's algorithm circuit used for factorising 15. All the symbols used in the diagram are presented in detail in Ref. \cite{Nielsen}. The system was initialised to the state $\ket{0000001}$ as described in \cite{Candela}. The first 3 qubits from top comprise the $x$-register and the next 4 qubits are the $f$-register. Part \textbf{(1)} corresponds to putting the $x$-register in a superposition of all possible values of $x$ from 0 to $2^3-1$. Part \textbf{(2)} corresponds to the gate-by-gate break down of the 3 function gates described in Ref. \cite{Candela}. These function gates control-multiply the $f$-register by $a^x$ mod 15. The resulting value of the $f$-register will be periodic. Part \textbf{(3)} represents the inverse Fourier transform which brings out the period of the function and we then read this out using the measurements on the far right of the diagram. The complete description of the process is available in Ref. \cite{Candela}\label{fig:shorsCircuit}.}
	\end{center}
\end{figure*}

An in-depth analysis of quantum algorithms can be found in the influential book of M. Nielsen and I. Chuang \cite{Nielsen}.

The qubit is the basic element of a quantum algorithm. It represents a linear superposition of two basis states usually denoted by $\ket{0}$ and $\ket{1}$ both of which can have complex coefficients. This representation is isomorphic to a 2D complex vector whose elements are given by the coefficients of the basis states. Then, naturally, the inner product between quantum states becomes a dot product between the vectors they represent. This trend continues to a system of an arbitrary number of qubits. As an important example, suppose we have a system comprising of 2 qubits. This system will be defined by a linear superposition of 4 basis states along with complex coefficients: $a\ket{00} + b\ket{01} + c\ket{10} + d\ket{11}$, where the first digit represents the state of the first qubit. This, in turn, can be represented as a 4D vector:

\begin{multline} a\ket{00} + b\ket{01} + c\ket{10} + d\ket{11}
\longleftrightarrow  \begin{pmatrix}
    a \\
    b \\
    c \\
    d \\
\end{pmatrix}.
\label{eq:2QubitSuperposition}
\end{multline}

We assume, without loss of generalisation, normalisation of all vectors.

Furthermore, it is a fact of quantum computation that all $n$ qubit gates with $n>3$ can be written in terms of 1 and 2-qubit gates \cite{Nielsen}\cite{Jones Logic Gates}. We carried out this decomposition on the entirety of Shor's algorithm for 7 qubits, as 1 and 2-qubit gates lend themselves well to our purpose. Figure \ref{fig:shorsCircuit} illustrates a simplified version of the circuit diagram given in Refs. \cite{Candela}\cite{Vandersypen}.

\section{Time-optimal evolution of pure states}\label{Sec:TimeOptimalSection}
One critical aspect of any type of computation, whether it is on a classical or quantum computer, is the computation time. This is the physical time it takes for a certain task to be performed. This section is devoted to describing the basics of time-optimal computation
with special regard to the implementation of Shor’s algorithm using NMR techniques by
employing the so-called quantum brachistochrone equation

\subsection{The Quantum Brachistochrone}

Quantum computation relies on the use of \textit{unitary operators} (as quantum gates) that map an input state to some output state depending on the nature of the gate employed. This is denoted, in the Schrödinger picture, as:

\begin{equation} \label{Unitary map}
\hat{U}(t_f, t_i) \ket{\psi (t_i)} = \ket{\psi (t_f)},
\end{equation}

\noindent where $\ket{\psi (t_i)}$ and $\ket{\psi (t_f)}$ denote the state at time $t_i$ initial and $t_f$ final respectively. We can observe that the nature of the time optimisation problem described above relies on time-optimising the transition between the initial input state $\ket{\psi (t_i)}$ to the final target output state $\ket{\psi (t_f)}$. We shall hence study \textit{what is the fastest possible way that any transformation of the form of (\ref{Unitary map}) can be done, subject to some initial constraints}. Such a task is already well described by what is known as the \textit{quantum brachistochrone equation}, which solved the following variational problem \cite{Carlini}\cite{Carlos}:
\begin{center} \label{variational formulation}
\textit{Given an initial pure state, $\ket{\psi_i}$, and a final state $\ket{\psi_f}$ find the possibly time-dependent Hamiltonian, $\hat{H}$, such that $\ket{\psi_i}$ evolves to $\ket{\psi_f}$ in the shortest possible time.}
\end{center}

The above problem was solved by proposing the action 

\begin{equation} \label{eq: Action}
 \begin{array}{l}
S(\psi, \phi, \hat{H}, \lambda) = \displaystyle\int dt \Bigl[ \  \frac{\sqrt[]{\bra{\dot{\psi}}(1-\hat{P}){\ket{\dot{\psi}}}}}{\Delta E} \ + \\ \mathrm{Re}  \left( i\braket{\dot{\phi}|\psi}+\bra{\phi}\hat{H}\ket{\psi} \right) 
+ {\sum}_m {\lambda}_m f_{m} (\hat{H}')\Bigr] , 
 \end{array}
\end{equation}
 
where $\hat{P}=\ket{\psi}\bra{\psi}$ represents the projector onto the state $\ket{\psi}$, $\Delta E$ is the energy variance of the Hamiltonian $\hat{H}$, $\ket{\phi}$, $\lambda$ are Lagrange multipliers and $f_m(\hat{H}')=0$ are the constraints imposed on  $\hat{H}'$, the traceless part of the Hamiltonian. The over-dot represents the time derivative. For convenience, we will from now on consider the Hamiltonian to be traceless as it can always be rescaled by:

\begin{equation} \label{make Vlad proud}
\hat{H}' = \hat{H}- \frac{Tr(\hat{H})}{N}\hat{\mathbb{1}},
\end{equation}

\noindent where $\hat{\mathbb{1}}$ is the identity and $N$ is the dimension of the Hilbert space. This is because we are more interested in the spacing of the energy levels (energy spectrum) rather than the actual value of the levels themselves \cite{Carlini}. From now onwards we will omit the dash notation employed in (\ref{make Vlad proud}) unless it is explicitly stated. \newline
	With this action \eqref{eq: Action} in hand, and by taking the variation with respect to $\ket{\phi}$, $\ket{\psi}$ and $\hat{H}$ we obtain

\begin{equation} \label{eq: what the F}
\hat{F} = \hat{F}\hat{P}+\hat{P}\hat{F}
\end{equation}

\noindent and the brachistochrone equation

\begin{equation} \label{eq: brachistochrone eq}
\frac{d}{dt}\hat{F} = i[\hat{H},\hat{F}]
\end{equation}
where 

\begin{equation} \label{eq: F this constraint}
\hat{F} \triangleq \sum_{m} \lambda_{m} \left( \frac{ \delta f_{m} }{ \delta \hat{H}} - \langle \frac{\delta f_m}{\delta \hat{H}}  \rangle \hat{P}  \right) .
\end{equation}

\noindent Given some constraints $f_m(\hat{H})$ we find $\hat{F}$ by using \eqref{eq: F this constraint}. With $\hat{F}$, we may now solve \eqref{eq: brachistochrone eq} with \eqref{eq: what the F} to obtain the time-optimal Hamiltonian $\hat{H}$ and the time-optimal state $\ket{\psi}$ that evolves from our initial state $\ket{\psi_i}$ to some final target state $\ket{\psi_f}$.

\subsection{Time-Optimal evolution}

An appropriate choice of constraint $f_m(\hat{H})$ is that that avoids the trivial infinitely energetic Hamiltonian that is indeed ``time-optimal" but far from physical. Thus, we avoid this problem by setting the variance of the energy spectrum to some constant $\omega$

\begin{equation} \label{general constraint}
\omega\\^2=\overline{E^2}-\overline{E}^2=\overline{E^2} \quad \because \ Tr(\hat{H})=0
\end{equation}

where $\overline{E^2}=\frac{Tr(\hat{H}^2)}{N}$ and $\overline{E}=\frac{Tr(\hat{H})}{N}$. In fact, it can be
mathematically shown that due to the nature of the variational problem

\begin{equation} \label{energy constraint}
\langle \hat{H} \rangle =\frac{Tr(\hat{H})}{N} \ \textrm{and} \ \langle \hat{H}^2 \rangle =\frac{Tr(\hat{H}^2)}{2},
\end{equation}

\noindent where $\langle \hat{H} \rangle$ denotes the expectation value of $\hat{H}$ with respect to $\ket{\psi}$ \cite{Carlini}. Expression (\ref{energy constraint}) implies

\begin{equation} \label{w is delte e. Still not impressed jonh?}
(\Delta E)^2 = \langle \hat{H}^2 \rangle-\langle \hat{H} \rangle^2=\langle \hat{H}^2 \rangle=\omega^2,
\end{equation}

\noindent where $ \Delta E$ is the energy variance and we have used expression (\ref{general constraint}). We are now in a position to assert the general mathematical form of our energy constraint,

\begin{equation} \label{DA CONSTRAINT}
f_m(\hat{H})=\frac{Tr(\hat{H}^2)}{2}-\omega^2=0.
\end{equation}

We can now obtain operator $\hat{F}$ from \eqref{eq: F this constraint} by using \eqref{DA CONSTRAINT}. With $\hat{F}$, we may now solve \eqref{eq: brachistochrone eq} with \eqref{eq: what the F} to obtain the time-optimal Hamiltonian $\hat{H}_{op}$, the time-optimal state $\ket{\psi}_{op}$ and the optimal time $T_{op}$\cite{Carlini}: 

\begin{equation} \label{eq: op ham (not a sandwitch though)}
\hat{H}_{op} = i\omega (\ket{\psi_f'}\bra{\psi_i}-\ket{\psi_i}\bra{\psi_f'}), 
\end{equation}

\begin{equation} \label{eq: op state}
\ket{\psi (t)}_{op} =\cos (\omega t) \ket{\psi_i} +\sin (\omega t) \ket{\psi_f'}, 
\end{equation}

\noindent and

\begin{equation} \label{eq: T-op}
T_{op} = \frac{1}{|\omega|} \arccos (|\braket{\psi_f|\psi_i}|),
\end{equation}

\noindent where $\ket{\psi_f'}$ and $\ket{\psi_i}$ denotes the final and initial state which are orthonormalised by using the Gram-Schmidt method.

\section{Experimental comparison using NMR quantum computation}\label{Sec:ExperimentalComparison}

    In 2001, a 7-qubit experimental realization of Shor's algorithm was accomplished by means of nuclear magnetic resonance (NMR) techniques \cite{Vandersypen}. These quantum computers work by initialising, controlling, manipulating and measuring the spins of nuclei within a cloud of molecules. These techniques are well documented in the literature \cite{Jones NMR,NMR1,NMR2,NMR3,NMR4,Levitt,Freeman}.
    
    A strong magnetic field is applied in the chosen $z$-direction across a collection of identical molecules. Under the influence of this field, the spin-$\frac{1}{2}$ nuclei within each molecule will either align or anti-align with the $z$-axis \cite{Shankar}. These are the two NMR qubit states.
   
    As the electron cloud varies from atom to atom, different atoms will have varying \textit{Larmor} frequencies, $\omega_0$,  within the same molecule \cite{NMR1}. This allows for the selective control of each nucleus \cite{Levitt}. An NMR quantum computer takes advantage of this to rotate individual qubit spin orientations about the orthonormal axes of the system, producing superpositions and effectively perform quantum gates. 
    
    Quantum spin transitions are induced by applying pulses of magnetic field rotating about the chosen $z$-axis. The rotating magnetic field is implemented by surrounding the cloud of nuclei with a coil connected to an AC voltage source \cite{Freeman}. This AC current will induce an effective magnetic field, which is proportional to the current passing through the coil. The Hamiltonian of a spin-$\frac{1}{2}$ nucleus in the lab frame under the influence of both the constant and rotating magnetic field becomes

	\begin{equation}
	\hat{H} = \omega_{0}\hat{I}_{z} + \omega_{n}[ cos(\omega_{r}t + \Phi)\hat{I}_{x} + \sin(\omega_{r}t + \Phi)\hat{I}_{y}],
    \label{eq:labframe_rf_hamiltonian}
	\end{equation}
	where $\omega_{n}$ represents the frequency at which the spin orientation nutates about the orthonormal axes, $\omega_{r}$ and $\Phi$ are the frequency and initial phase of the AC circuit respectively and the $\hat{I}$ terms represent the Pauli spin matrix notation used by Jones et. al. \cite{Jones NMR}. This Hamiltonian in the $\omega_{r}$ rotating frame becomes,
	\begin{equation}
	\tilde{H} = [\omega_{0} - \omega_{r}]\hat{I}_{z} + \omega_{n}[\cos(\Phi)\hat{I}_{x} + \sin(\Phi)\hat{I}_{y}].
	\label{eq:rotating_rf_hamiltonian}
	\end{equation}
	%
	By setting the frequency of the AC circuit such that $\omega_{0}=\omega_{r}$, the $z$-spin component of the Hamiltonian will cancel. Thus, the quantisation in the stationary frame is maintained along the $z$-axis. Regardless, the rotating magnetic field pulses cause the nuclear spin to rotate about a different axis, hence putting it into a superposition of the qubit states. 
	
	We are now able to control the phase $\Phi$ of our AC circuit and thus select the axis of rotation \cite{Jones}. By combining rotations about the $x$ and $y$-axes, we can effectively induce any single qubit quantum gate on our system.

	If we now wish to perform a single qubit operation, such as a NOT gate, we induce a magnetic field in the $x$-direction by setting $\Phi=0$. This operation would thus have the following Hamiltonian:
	
	\begin{equation}
	\tilde{H}_{\text{NOT}} =  \omega_{n} \hat{I}_{x}.
	\label{eq:HamiltoanX}
	\end{equation}
	Any qubit state $\ket{\tilde{\psi}}$, under the action of this Hamiltonian, will evolve over time like
	\begin{equation}
	\ket{\tilde{\psi} (t)} = \exp{(-i\omega_{n} \hat{I}_{x} t)}\ket{\tilde{\psi}(0)}.
	\label{eq:psitX}
	\end{equation}
	After a time of $\text{t}=\frac{\pi}{\omega_{\text{n}}}$ the state will evolve such that the exponential term in \eqref{eq:psitX} becomes $-i\hat{I}_{x}$. This matrix corresponds to the NOT gate operation up to a global phase of $-i$. Fortunately, this can be neglected \cite{Jones NMR}.

	We denote this single NOT rotation as $180^{\circ}_{x}$ as it is applied by inducing a rotation of $180^{\circ}$ about the $x$-axis. Likewise any single rotation of an angle $\phi$ about either the $x$ or $y$-axis will take an evolution time of:
    
    \begin{equation}
	t_{\phi} = \frac{\phi}{\omega_{n}} 
	\label{eq:Phi_evolution_time}.
	\end{equation}
Thus, if we apply the same pulse but for half the time span, we end up with the operation of a square-root NOT gate which is also used in the simulation. The Hadamard gate is experimentally implemented using a sequence of three pulses:
	\begin{equation}
	45^{\circ}_{y} \rightarrow 180^{\circ}_{x} \rightarrow 45^{\circ}_{-y}.
	\label{eq:Hadamard_Pulse_Sequence}
	\end{equation}
    This gate is implemented such that it always acts as a self-inverse operation regardless of the initial qubit state \cite{Jones NMR}.
    
    We have therefore introduced how gates are implemented for single qubits. An additional level of complexity is added when implementing controlled gates. Let us consider the CNOT gate. This operation is applied by sandwiching a controlled $\hat{\pi}$ phase gate between two Hadamard gates acting on the controlled qubit \cite{Jones Logic Gates}.

	The matrix representation of this gate can be expressed as the sum of the two-qubit spin operators defined in Jones et. al. \cite{Jones NMR}:
	\begin{equation}
	\hat{\pi} = \exp[\pm i\frac{\pi}{2}(\frac{1}{2}\hat{\mathbb{1}} - \hat{I}_{z} - \hat{S}_{z} + 2\hat{I}_{z}\hat{S}_{z})],
	\label{eq:pi_gate_equation}
	\end{equation}
	with the $\frac{1}{2}\hat{\mathbb{1}}$ term included as a global phase shift which can be neglected. The coupled term ($2\hat{I}_{z}\hat{S}_{z}$) is implemented using a spin-echo technique \cite{Freeman} which involves a sequence of $x$ and $y$-axis rotations acting on \textit{both} qubits. The $\hat{I}_{z}$ and $\hat{S}_{z}$ operations are implemented by periods of free precession about the $z$-axis which can be interpreted as rotating the $\omega_{r}$ reference frame. Due to the fact that operations on different qubits commute, they can be applied in any order, thus a variety of combinations can be implemented. Some are favoured over others as rotations are cancelled. One such optimal combination, given in the literature \cite{Jones NMR}, is
	\begin{equation}
	\frac{1}{4J} \rightarrow \space 180^{\circ}_{x} \rightarrow \frac{1}{4J} \rightarrow 90^{\circ}_{x} \rightarrow 90^{\circ}_{-y} \rightarrow 90^{\circ}_{x},
	\label{eq:Pi_phase-gate-implementation}
	\end{equation}
	with each pulse acting on both spins and the $\frac{1}{4J}$ representing the periods of free evolution given by the spin-coupling frequency $J$ which is intrinsic to any pair of two spin-$\frac{1}{2}$ nuclei within the same molecule. 
	
	To further simplify the CNOT operation, we can replace the Hadamard gates on either side of the phase gate with an inverse pseudo-Hadamard ($90^{\circ}_{-y}$) and a pseudo-Hadamard ($90^{\circ}_{y}$). Alone these operations are non-unitary and are not good substitutes for the Hadamard operation given in equation \eqref{eq:Hadamard_Pulse_Sequence}, but when combined with the phase gate rotations \eqref{eq:Pi_phase-gate-implementation} form a unitary CNOT operation.
	
	The controlled square-root NOT and $\phi=\frac{\pi}{4}$ phase gates are performed in a similar manner to the controlled NOT gate, the main difference being that the rotation time given for some operation steps are shortened. Implementation methods for these operations are explained in more detail in the literature \cite{Jones}.

    By adding up the time to make each rotation induce these phase gates using equation \eqref{eq:Phi_evolution_time} along with the duration of the $J$-coupling intervals, we get the following expressions for the time taken to complete the phase gate operation:
\begin{equation}
\begin{split} 
t^{\ 180^{\circ}} = \frac{3\pi}{2\omega_{n}} + \frac{1}{2J}, & \qquad t^{\ 90^{\circ}} = \frac{9\pi}{4\omega_{n}} + \frac{1}{4J}, \\ \qquad t^{\ 45^{\circ}} = & \frac{11\pi}{8\omega_{n}} + \frac{1}{8J},
\label{eq:t_phase}
\end{split}
\end{equation}
 where the superscript denotes the angle of the phase gate.
    
    We now have all the tools needed to calculate the time it would take to physically implement this version of Shor's algorithm using NMR techniques.

    \subsection{Experimental Duration of Shor's Algorithm in NMR}

    \quad Section \ref{Sec:ExperimentalComparison} portraits how the gate operations in Shor's algorithm can be implemented using NMR techniques. Moreover, we introduced a way of calculating the time it takes to perform each operation using equations \eqref{eq:Phi_evolution_time} and \eqref{eq:t_phase}. Employing these, we only need to know the $\omega_{n}$ and $J$-coupling values used in the experimental realisation \cite{Vandersypen}. 
    
    Although the $J$-coupling values are provided, the exact values needed to calculate $\omega_{n}$ had to be estimated. We can use the approximate times given for all single qubit rotation to acquire a good estimate for $\omega_{n}$. This approximation is made because we are told the pulse times range from `$0.22$ to ${\sim}2\text{ms}$' \cite{Vandersypen}. As our shortest rotations $(22.5^{\circ}_{-y})$ take 8 times less time than the $(180^{\circ}_{-y})$ rotations then it implies that $t_{\phi}= \frac{\pi}{\phi} \times 1.88 \pm 0.12$. Therefore by using equation \eqref{eq:Phi_evolution_time} we estimate $\omega_{n} = 1670 \pm 110 \text{H}_{z}$ (further experimental data would aid testing the theory proposed). This is a reasonable estimation as nutation frequencies tend to be in the low radio-frequency range \cite{Freeman}. Using this $\omega_{n}$ along with the $J$-coupling values of our system from the paper by Vandersypen et. al. \cite{Vandersypen}, we can plot the cumulative time of each of the 25 gates (see Figure \ref{fig:finalcomparison}).
    
We calculate the total time an NMR quantum computer would take to perform the quantum part of Shor's algorithm to be $1.59 \pm 0.04 \text{s}$. In particular, the time taken to perform the inverse quantum Fourier transform part (steps 20-25) of our computation is $112 \pm 2 \text{ms}$, as portrayed in Figure \ref{fig:finalcomparison}. We cannot compare the total time with that given in the experimental realisation due to certain gates being “removed” and “replaced by simpler gates” \cite{Vandersypen}, yet no information is given on what these adaptations are. However, the inverse quantum Fourier transform is performed in the same way as in our simulation and is quoted as taking of the order of $120 \text{ms}$ to implement which compares well to our theoretical value. Discrepancies between the times are likely due to the nuclei representing the first and third qubits having similar gyromagnetic frequencies, thus a few short pulses were added in the experiment to refocus one nucleus after the other has been operated on. For simplicity, we have omitted the times of the refocusing pulses as they do not take a significant amount of time \cite{Freeman}. As errors are not given in the data-set \cite{Vandersypen}, we assume the errors are given by the precision of the values provided.

\section{Calculating the time-efficiency of the experiment} \label{Sec:TimeEfficiencyOfExperiment}

Calculating the expected optimal time presents the problem that it cannot be approached analytically as one cannot predict the quantum state of the system due to the stochastic nature of the problem. However, one can bypass such issue by performing a numerical simulation with an arbitrarily large number of states (1 million states in our case). In this way, one could at least predict what the average expected time duration would be if an experiment was to be performed. Hence, we programmed a simulation with the circuit described in Figure \ref{fig:shorsCircuit} and with the time-optimal gates described in Section \ref{Sec:TimeOptimalSection}. \newline
\indent The simulation selected a random point on a unit 2D circle (or 4D sphere in the case of 2-qubit gates) as the input state, took the final vector according to the effect of a given gate on the input state and calculated the optimal time according to equation \eqref{eq: T-op}. This process was repeated with different random initial states and averaged over 1 million times to get the expected optimal time for the given gate. In the simulation, $\omega$ denoted the physical parameters which were calculated in Section \ref{Sec:ExperimentalComparison}. \newline
\indent Finally, all expected times were added together to obtain the overall expected time of the experiment. This proposed method is a fast way to check the time-efficiency of a given quantum experiment, where time is a crucial asset, such as in the computational field.

\section{Results \& Conclusion}\label{Sec:Results & Conclusion}

\begin{figure}[h!]
	\begin{center}
	\includegraphics[scale=0.2]{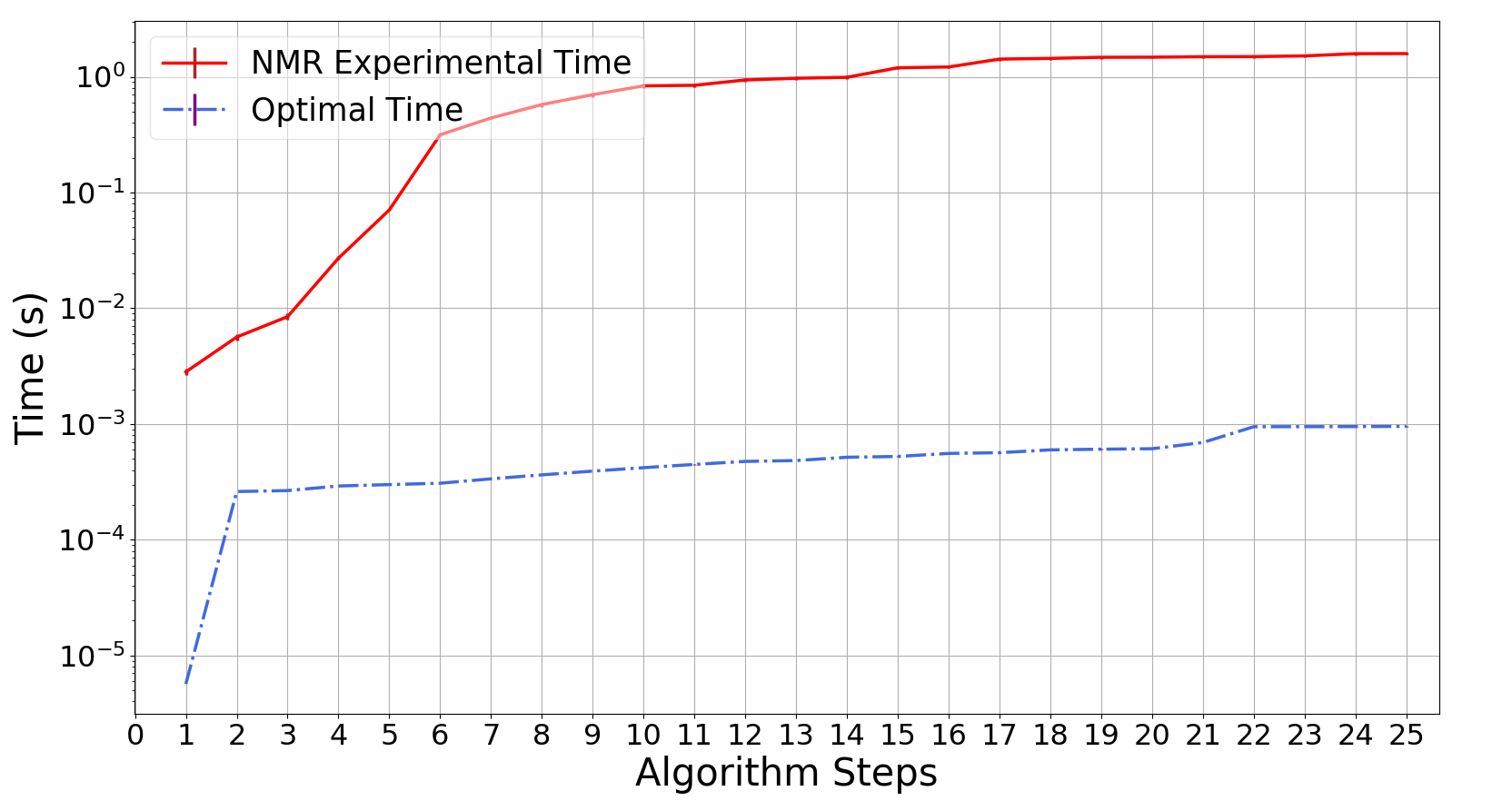}
	\caption{\label{fig:finalcomparison}Plot comparing the duration of Shor's algorithm for factorising 15 using NMR techniques and time-optimal translations. The scale is logarithmic such that both data sets appear clearly on the plot. The $x$-axis represents each of the 25 steps of the quantum computation given in Figure \ref{fig:shorsCircuit}. The $y$-axis shows the cumulative time it takes to perform each step in our algorithm, calculated using the frequency values given by Vandersypen et. al. \cite{Vandersypen}. Errors are not quoted but we assume the error is given by the precision of the values provided, although they are not clearly visible on the logarithmic scale.  The 0 point on the $x$-axis is omitted as it corresponds to $\log(0)=-\infty$ on the logarithmic scale. This plot was made using Python.}
	\end{center}
\end{figure}

In order to meaningfully present the data, we plotted a logarithmic scale of the expected optimal time of the circuit and we also included the calculated time of the experiment. Findings are shown in figure \ref{fig:finalcomparison}.

The key point is that there is a 3 order of magnitude difference in the time taken by the experiment compared to that obtained using the variational approach. Thus, we show that there is significant room for improvement in terms of duration of this algorithm. However, our method does not take into account technical difficulties which the implementation might present. Therefore, the result is to be interpreted rather as theoretical lower bound of time for the NMR implementation of Shor's algorithm for factorising 15. 

Another important aspect of our work is that the method can be extended to any scheme for implementing quantum algorithms, since all that is needed are the experimental parameters $\omega$ and since all gates which act on 3 or more qubits can be broken down into 1 and 2 qubit gates.

\section{Further Work}
The natural continuation of this work would be to extend the above methodology to mixed states as opposed to pure quantum states. This continuation could be based on Ref. \cite{MixedStates} and implement a similar Monte Carlo approach for calculating the expected optimal time as described previously. The proposed work could be more suitable for instance in cases where quantum decoherence plays a big role in the experiment.

\section*{Acknowledgements}

We would like to thank our project supervisor, Dr. Edward McCann, our lecturers Dr. Henning Schomerus and Dr. Alessandro Romito for capital guidance throughout the whole process and our colleague Niall Mulholland for his support in the organisation surrounding the project.

\pagebreak

\end{document}